\documentclass[prl,superscriptaddress,twocolumn,showpacs,preprintnumbers,amsmath,amssymb]{revtex4}
\usepackage{hyperref}
\usepackage{epsfig}
\usepackage{multirow}
\usepackage{xspace}
\usepackage{atlasphysics}

\newcommand{\bone}   {\tilde{b}^{}_{1}}

\newcommand{\neut}  {\tilde{\chi}^{0}_{1}}  
 
\def\etmiss{\ensuremath{E_{\mathrm{T}}^{\mathrm{miss}}}\xspace}

\def\dphimin{\ensuremath{\Delta\phi_{\mathrm{min}}}}

\def\bjets{$b$-jets\xspace}

\def\btagging{$b$-tagging\xspace}

\def\bquark{$b$-quark\xspace}
\def\bquarks{$b$-quarks\xspace}

\def\Wjets{$W$+jets\xspace}
\def\Zjets{$Z$+jets\xspace}

\def\WHF{$W$+hf\xspace}
\def\ZHF{$Z$+hf\xspace}
\renewcommand{\ttbar} {\ensuremath{t\bar{t}}\xspace}
\newcommand{\ttbb}{\ensuremath{t\bar{t}b\bar{b}}\xspace}
\renewcommand{\met} {\ensuremath{E_{\mathrm{T}}^{\mathrm{miss}}}\xspace}
\renewcommand{\pt} {\ensuremath{p_\mathrm{T}}\xspace}
\newcommand{\mct}{\ensuremath{m_\mathrm{CT}}}
\newcommand{\lumi}{2.05}
\newcommand\totallumi{2.05~${\rm fb}^{-1}$}

\def\met{\ensuremath{E_{\mathrm{T}}^{\mathrm{miss}}}}

\begin{document}
\preprint{CERN-PH-EP-2011-195}
\preprint{Submitted to Physical Review Letters}
\vspace*{0.2cm}
\title{Search for Scalar Bottom Quark Pair Production with the ATLAS Detector \\ in $pp$ Collisions at $\sqrt{s}=$~7~TeV}
\author{The ATLAS Collaboration} \affiliation{}
\begin{abstract} 
The results of a search for pair production of the scalar partners of bottom quarks 
in \totallumi~of $pp$ collisions at $\sqrt{s}=7$~TeV using the ATLAS experiment are 
reported. Scalar bottom quarks are searched for in events with large 
missing transverse momentum and two jets in the final state, where both jets are 
identified as originating from a bottom quark. In an $R$-parity conserving minimal 
supersymmetric scenario, assuming that the scalar bottom quark decays exclusively 
into a bottom quark and a neutralino, 95\% confidence-level upper limits are obtained in 
the $\bone-\neut$ mass plane such that for neutralino masses below 60~GeV scalar 
bottom masses up to 390~GeV are excluded.

\end{abstract}
\maketitle
\normalsize

%%% introductory matter
Supersymmetry 
(SUSY)~\cite{Golfand:1971iw,Neveu:1971rx,Neveu:1971iv,Ramond:1971gb,Volkov:1973ix,Wess:1973kz,Wess:1974tw,Fayet:1977yc,Farrar:1978xj}  
is a theory that provides an 
extension of the Standard Model (SM) and naturally resolves the hierarchy 
problem by introducing supersymmetric partners of the known bosons and 
fermions. In the framework of a generic $R$-parity conserving minimal 
supersymmetric extension of the SM, the MSSM, SUSY 
particles are produced in pairs and the lightest supersymmetric particle 
(LSP) is stable, providing a possible candidate for dark matter. 
%In the MSSM, 
The scalar partners of right-handed and left-handed quarks (squarks) 
%, $\sqR$ and $\sqL$, 
can mix to form two mass eigenstates. 
In a large variety of models, the LSP is the lightest neutralino ($\neut$) and 
the mass of the lightest scalar bottom quark eigenstate (sbottom, $\bone$) can 
be significantly lower than the other squark masses.  
%The mixing is proportional to the mass of the corresponding SM fermion 
%and therefore becomes important for the third generation. Large mixing 
%can yield sbottom ($\bone$) and stop ($\tone$) mass eigenstates which are 
%significantly lighter than other squarks. 
Consequently, $\bone$ could be produced with large cross sections at the Large Hadron Collider (LHC).
In this Letter, a search for direct sbottom pair production is presented, 
assuming a SUSY particle mass hierarchy such that the sbottom decays exclusively 
into a $b$-quark and a $\neut$ ($\bone \rightarrow b\neut$). 
The expected signal is characterized by the presence of two jets originating 
from the hadronization of the \bquarks and missing transverse momentum --- its 
magnitude is referred to as $\etmiss$ in the following --- resulting from the 
undetected neutralinos. Results of searches for direct sbottom 
production have been previously reported by the Tevatron~\cite{sbottomCDF,sbottomD0} and LEP~\cite{lep2} experiments.  
The search described here extends these results using 
\totallumi of 7~TeV $pp$ collision data recorded by the 
ATLAS experiment at the LHC. 

%%% ATLAS
The ATLAS detector~\cite{DetectorPaper:2008} consists of inner tracking devices
surrounded by a superconducting solenoid, electromagnetic
and hadronic calorimeters, and a muon spectrometer
with a toroidal magnetic field. The inner detector,
in combination with the 2~T field from the solenoid,
provides precision tracking of charged particles for $|\eta| < 2.5$~\cite{atlascoordinates}.
 It consists of a silicon pixel detector, a silicon strip
detector, and a straw tube tracker that also provides transition
radiation measurements for electron identification.
The calorimeter system covers the pseudorapidity range
$|\eta| < 4.9$. It is composed of sampling calorimeters with
either liquid argon (LAr) or scintillating tiles as the active
media. The muon spectrometer has separate trigger and high-precision 
tracking chambers which provide muon identification and measurement 
for $|\eta| < 2.7$.

%%% Object selection
This search uses data recorded between March and August 2011 at the LHC. 
After the application of beam, detector, and data quality requirements, the data 
set corresponds to a total integrated luminosity of 
2.05$\pm$0.08~fb$^{-1}$~\cite{Aad:2011dr,ATLAS:Lumi}. 
The data are selected with a three-level trigger system that requires a high transverse momentum (\pt) jet 
and  missing transverse momentum~\cite{trigger}. Events are required to have a reconstructed primary 
vertex associated with five or more tracks, consistent with the beam spot position. 

Jets are reconstructed offline from three-dimensional topological calorimeter 
energy clusters by using the anti-$k_t$ jet algorithm~\cite{Cacciari:2008gp,Cacciari200657} 
with a radius parameter of 0.4. The measured jet energy is corrected 
for inhomogeneities and for the non-compensating nature of the calorimeter 
by using $\pt$- and $\eta$-dependent correction factors (\cite{Aad:2011he} and 
references therein). Only jet candidates with $\pt > 20$~GeV 
within $|\eta| < 2.8$ are retained. 
During the data-taking period, a localized electronics failure in the 
LAr barrel calorimeter created an inactive region in the second and third 
calorimeter 
layers ($\Delta \eta \times \Delta \phi \simeq 1.4 \times 0.2 $) in which on average 
30\% of the incident jet energy is not measured. 
If either of the two selected 
leading jets fall in this region the event is rejected. 
The loss in signal acceptance is smaller than 10\% for the models considered. 

Electron candidates are required to have $\pt > 20$~GeV, $|\eta| < 2.47$,  
and satisfy the `medium' selection criteria reported in~\cite{elperf}. Muon 
candidates are required to have $\pt > 10$~GeV, $|\eta| < 2.4$ and are 
identified as a match between an extrapolated inner detector track and one or more 
track segments in the muon spectrometer. 
Additional requirements are applied to electron and muon candidates when 
defining lepton ($\ell$=$e, \mu$) control regions. In this case, electrons must have 
$\pt > 25$~GeV, pass the `tight' selection criteria in~\cite{elperf} and  
be isolated --- the \pt sum of tracks within a cone in the $(\eta, \phi)$ plane of radius 
$\Delta R = 0.2$ around the candidate, $\Sigma \pt$, must be less than 
10\% of the electron \pt. 
Muons must have $\pt > 25$~GeV, longitudinal and transverse impact parameter within 1~mm and 0.2~mm 
of the primary vertex, respectively, and $\Sigma \pt< 1.8$~GeV. 
Following the object reconstruction described above, overlaps between jet candidates and 
electrons or muons are resolved. Any jet within a distance $\Delta R $ = 0.2 of 
a `medium' electron candidate is discarded. 
Any remaining lepton within $\Delta R=$ 0.4 of a jet is discarded. 

The calculation of $\etmiss$ is based on the magnitude of the vectorial sum 
of the $\pt$ of the reconstructed jets (with $\pt > 20 \GeV$ and $|\eta| < 4.5$), 
leptons --- including non-isolated muons --- and the calorimeter clusters 
not belonging to reconstructed objects~\cite{ATLAS-MET}.

Events must pass basic quality criteria against 
detector noise and non-collision backgrounds and 
are selected with at least one jet with $\pt>130$~GeV
and  $\etmiss > 130$~GeV to ensure full trigger efficiency, and one
additional jet with $\pt > 50$~GeV. 
Both jets are required to be associated to the primary interaction and to originate 
from a \bquark using a tagging algorithm that exploits both impact parameter 
and secondary vertex information. Jets are tagged for $|\eta|<2.5$ and   
the parameters of the algorithm are chosen such that a tagging efficiency 
of 60\%, 10\% and $<$1\% is achieved for $b$-jets, $c$-jets, and light flavor or 
gluon jets, respectively, in $\ttbar$ events in Monte Carlo (MC) simulation~\cite{ATLAS:BtagCalib}. 
Events with identified electron or muon candidates or additional 
jets with $\pt > 50$~GeV are vetoed to exploit 
the topology of the addressed signal.  
The multijet background contribution with large \met, due to the 
mismeasurement of the jet energies in the calorimeters or to neutrino 
production in heavy quark decays, is suppressed by requiring the smallest 
azimuthal separation between the $\etmiss$ direction 
and any of the two leading jets, $\dphimin$, to be above 0.4. If a third jet 
with $30<\pt<50$~GeV is found in the event, its azimuthal separation 
with the \met~must be above 0.2, where jets with \pt below 30~GeV are 
not used due to the large expected contribution from multiple interactions.  
In addition, the ratio of \met~to the scalar sum of \met~and the 
transverse momenta of the two leading jets is required to be above 0.25.   

A selection based on the boost-corrected contransverse mass~\cite{Polesello:2009rn}, \mct,  
is employed to further discriminate  sbottom pair production from SM background processes.  
For two identical decays of heavy particles into two visible particles $v_{1}$ 
and $v_{2}$, and into invisible particles, the contransverse mass~\cite{Tovey:2008ui} 
is defined as 
$\left([ E_{\rm T}(v_{1}) + E_{\rm T}(v_{2}) \right]^2 -  \left[  {\bf p_T}(v_{1}) - {\bf p_T}(v_{2}) \right]^2)^{1/2}$. 
The boost-corrected contransverse mass conservatively corrects to account for boosts in the 
transverse plane due to initial state radiation  and hence protects the 
expected endpoint in the distribution. 
In the case of sbottom pair production with $\bone \rightarrow b\neut$, \mct~is 
expected to have an endpoint at ($m(\bone)^2-m(\neut)^2)/m(\bone)$. 
To maximize the sensitivity 
across the $\bone-\neut$ mass plane, three signal regions (SR) are defined as 
a function of the \mct~threshold, with \mct$ \, > 100$, 150 and 200~GeV. 

%%% Background estimate
Simulated event samples are used to aid in the description of the background and 
to model the SUSY signal. 
Top quark pair and single top production are simulated with {\tt MC@NLO}~\cite{mcatnlo}, 
fixing the top quark mass at 172.5~GeV, and the next-to-leading-order (NLO) parton density function (PDF) set 
{\tt CTEQ6.6}~\cite{cteq66}. Samples of \Wjets, \Zjets with light 
and heavy flavor jets, and \ttbar with additional \bjets, \ttbb, are generated with 
{\tt ALPGEN}~\cite{alpgen} and PDF set {\tt CTEQ6L1}~\cite{cteq6l1}.  
The fragmentation and hadronization for the {\tt ALPGEN} and {\tt MC@NLO} samples are 
performed with {\tt HERWIG}~\cite{herwig}, using {\tt JIMMY}~\cite{jimmy} for the 
underlying event. 
Samples of $Z \ttbar$ and $W \ttbar$ are generated with {\tt MADGRAPH}~\cite{madgraph} 
interfaced to {\tt PYTHIA}~\cite{pythia}. Diboson ($WW$, $WZ$, $ZZ$) samples 
are generated with {\tt HERWIG}. 
All background cross sections are normalised to the results of higher 
order calculations~\cite{Aad2011398}.  
The signal samples are generated in the MSSM framework at fixed $\bone$ 
and $\neut$ masses, with BR($\bone \rightarrow b\neut$)=100\%, using the
{\tt HERWIG++}~\cite{herwig} v2.4.2 Monte Carlo program. 
The SUSY sample yields are normalized to the results of NLO calculations, as obtained using the {\tt PROSPINO}~\cite{Beenakker:1997ut} 
v2.1 program. The {\tt CTEQ6.6M}~\cite{cteq6m} parameterization of the PDFs is used and the renormalization and factorization scales, 
$\mu^{}_{\mathrm R,\mathrm F}=\mu$,  are set to the sbottom mass.  
NLO cross sections vary between 12~pb and 0.05~pb for sbottom masses 
in the range 200-500~GeV. 
The MC samples are produced using parameters tuned as described in~\cite{atlastune1,atlastune2} and are 
processed through a detector simulation~\cite{atlassimulation} based on {\tt GEANT4}~\cite{geant4}. 
Effects of multiple proton-proton interactions per bunch crossing are included in the simulation. 

The signal efficiencies vary across the $\bone$-$\neut$ mass plane.   
As an example, for the SR with \mct$\, > 150$~GeV the efficiencies are found to 
be between 1\% and 6\% as the sbottom mass increases from 200~GeV to 500~GeV, and 
between 6\% and 2\% as $\Delta m=m^{}_{\bone}- m^{}_{\neut}$ decreases. 
The sensitivity for $\Delta m$ below 130~GeV is limited by the selection in \mct~and 
the trigger-driven requirements on the leading jet \pt and \met.

The main SM processes contributing to the background are top quark pair and single top 
production as well as associated production of $W/Z$ bosons with heavy flavor 
jets --- referred to as \ZHF and \WHF, respectively. In particular, the signal region with \mct$\, > \, 100$~GeV is 
dominated by semi-leptonic \ttbar events as a consequence of the \mct~endpoint 
at 135~GeV in top pair production, whilst  the signal regions with \mct$\, > \, 150, 200$~GeV are dominated 
by the irreducible \ZHF~production with $Z \rightarrow \nu \nu$ decay, followed 
by \WHF~production with $W \rightarrow \tau \nu$.   
Contributions from multijet, diboson and associated production of \ttbar with 
$W,\ Z$ or additional \bjets are sub-dominant. 
Non-collision backgrounds were found to be negligible. 

The estimation of the main background processes is carried out by defining 
data control regions where each background component is dominant.
The background estimate in each SR is derived through 
`transfer factors' equivalent to the ratio of expected event yields in the 
signal and control regions estimated using the MC. 
The contributions from top and \WHF~production are estimated using a 
transfer factor from a control region where events have exactly one electron or 
muon, \etmiss$\, >$~80~GeV, and at least two \bjets~with \pt$\, >$~130~GeV and 
50~GeV. The transverse mass of the ($\ell, \etmiss$) 
system is required to be between 40~GeV and 100~GeV to select 
events containing  $W\rightarrow \ell \nu$. The contribution to this control region from other SM processes 
accounts for less than 10\% of 
the total and is estimated from simulation. 
The \ZHF~contribution is estimated using a transfer factor from a control region 
where events have two opposite-sign same-flavor leptons 
($\ell^+ \ell^-$), at least two \bjets with \pt$\, >$~80~GeV and 50~GeV, and 
invariant mass of the two leptons $m_{\ell \ell}$ between 81~GeV and 101~GeV 
($Z$-mass interval). In addition, the transverse momentum of the 
($\ell^+,\ell^-, ~\etmiss$) system is required to be greater than 50~GeV. 
%The momenta of the leptons are added to the \etmiss~to mimic the $Z\rightarrow \nu \nu$ decay. 
The contribution from top quark production in 
this control region accounts for about 50\% of the total and is subtracted using a side-band estimate 
in two 40~GeV mass windows above and below the $Z$-mass interval. 

The sub-dominant background contribution from dibosons, 
Z\ttbar, W\ttbar and \ttbb~is estimated using MC simulation and 
increases from 1\% to 10\% of the total SM prediction as the 
selection cut on \mct~increases from 100~GeV to 200~GeV. 
Finally, the residual multijet background is estimated using data. 
A sample of multijet events with large \etmiss~is constructed starting from 
multijet events with low \etmiss~and smearing the momenta of jets with 
response functions. This prediction is tuned in a multijet dominated control 
region where the requirement on \dphimin~is inverted~\cite{0leptonPLB}. 
The contribution is found to be less than 5\% across the SRs. 
 
%%% Systemaics
The total systematic uncertainty on the background expectations varies from 21\% to 44\%, 
increasing with the \mct~selection applied, and is dominated by the  uncertainty due 
to finite data statistics in the control regions. 
The next dominant uncertainty on the 
SM estimates derives from the residual uncertainties on the theoretical 
modeling of the top~background. It varies between 10\% and 15\% depending 
on the SR and is evaluated using additional MC samples: {\tt ACERMC}~\cite{Kersevan:2004yg},~\cite{topcross}  
for the impact of initial and final state radiation, %parameters were varied
{\tt PYTHIA} and {\tt POWHEG}~\cite{powheg} for choice of fragmentation 
model and generator.  
%an alternative fragmentation model ({\tt PYTHIA}) and an 
%alternative generator ({\tt POWHEG}~\cite{powheg}). 
The residual uncertainties on the \WHF~and~\ZHF theoretical modeling account for less than 5\% of the total 
uncertainty.  
Finally, the experimental uncertainties on the \btagging~efficiency and jet energy 
scale and resolution are also considered; across the SRs 
they vary between 5\% and 8\% and between 6\% and 9\%, respectively.   

For the SUSY signal processes, uncertainties on renormalization and 
factorization scales and on the PDF affect the theoretical cross section. 
PDF uncertainties are estimated using the {\tt CTEQ6.6M} PDF error 
eigenvector set 
and are between 7\% and 16\% depending on the sbottom mass.  
The variation of renormalization and factorization scales by a factor 
of two changes the nominal signal cross section, $\sigma^{}_{\rm{nom}}$, by $\pm$15\%, 
independently of $m^{}_{\bone}$. 
In the following, the cross sections calculated with scale settings 
2$\times \mu$ and $\mu/2$ are referred to as $\sigma^{}_{\rm{min}}$ and 
$\sigma^{}_{\rm{max}}$, respectively. 
The impact of detector-related uncertainties, such as the jet energy 
scale and \btagging, on the signal event yields varies between 35\% 
and 45\% and is dominated by the uncertainties on the \btagging efficiency.  

\begin{figure}[!h]
\centering                                                            
 \includegraphics[width=1.0\columnwidth]{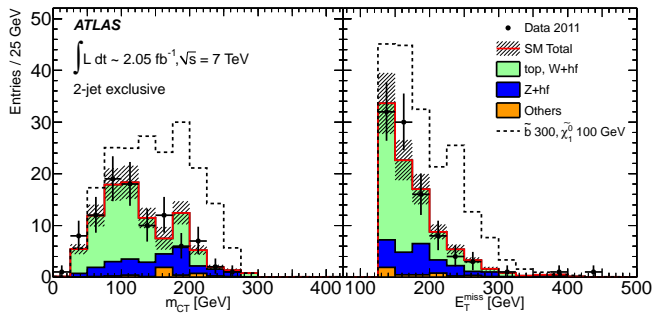}
\caption{Measured \mct~(left) and \met~(right) distributions before the 
\mct~selection 
compared to the SM predictions 
(solid line) and SM+MSSM predictions (dashed lines). 
The dashed grey band represents the total systematic
uncertainties.}\label{fig:res_plots}
\end{figure}

%%% results and limits
Table~\ref{tab:0lepFinalResultsSR_2jX} reports the observed number of events 
and the SM predictions before the \mct~selection and for each SR. Both transfer factor and MC estimates are given. The data are 
in good agreement with the SM background expectations within uncertainties 
in all cases.   
Figure~\ref{fig:res_plots} shows the measured \mct~and \met~distributions 
before \mct~selection compared to the SM predictions.  
MC estimates are used, rescaled to match the total integral 
and systematic uncertainties predicted by the transfer factor 
estimates for \ZHF and the sum of top and \WHF, respectively. For illustrative 
purposes, the distributions expected for the MSSM scenario with  
sbottom and neutralino masses of 300~GeV and 100~GeV, respectively, are added 
to the SM predictions. 
\begin{table}[htp]
    \renewcommand\arraystretch{1.1}
 \centering
 \begin{tabular}{l|ccc|c|c}
   \hline\hline
   \mct  & top, \WHF & \ZHF & Others & Total SM & Data \\
    GeV & TF & TF & MC+JS &  &  \\
        & (MC) & (MC) &  &  &  \\
   \hline
$ 0 $   & 67 $\pm$ 10     & 23 $\pm$ 8   & \multirow{2}{*}{3.6 $\pm$ 1.5} & 94 $\pm$ 16 & 96 \\
        & (60 $\pm$ 25)   & (16 $\pm$ 9) &             & (80 $\pm$ 35) &  \\
$ 100 $ & 36 $\pm$ 10      & 23 $\pm$ 9   & \multirow{2}{*}{3.1 $\pm$ 1.6} & 62 $\pm$ 13 & 56 \\
        & (34 $\pm$ 16)   & (12 $\pm$ 7) &             & (49 $\pm$ 25) &  \\
$ 150 $ & 12 $\pm$ 5      & 12 $\pm$ 6   & \multirow{2}{*}{2.7 $\pm$ 0.9} & 27 $\pm$ 8 & 28 \\
        & (13 $\pm$ 8)    & (8.3 $\pm$ 4.7)  &             & (24 $\pm$ 13) &  \\
$ 200 $ & 3.2  $\pm$ 1.6      & 3.9  $\pm$ 3.2   & \multirow{2}{*}{1.0 $\pm$ 0.9} &  8.1 $\pm$ 3.5  & 10 \\
        & (4.1 $\pm$ 3.4)     & (2.8 $\pm$ 1.5)  &             &  (8.0 $\pm$ 4.9) &  \\
  \hline\hline
  \end{tabular}
  \caption{Expected and measured number of events for an integrated luminosity of \lumi~fb$^{-1}$. 
  \ZHF and the sum of top and \WHF~are estimated using a transfer factor estimate (TF). 
  The column labelled as `Others' reports multijet background predictions as estimated with a 
  jet smearing (JS) method, and sub-dominant SM backgrounds estimated from MC.    
  For comparison the numbers obtained using MC samples are shown in parenthesis. 
  The total systematic uncertainties are also reported. } 
  \label{tab:0lepFinalResultsSR_2jX}
\end{table}
\begin{figure}[h!]
  \begin{center}
      \includegraphics[width=0.95\columnwidth]{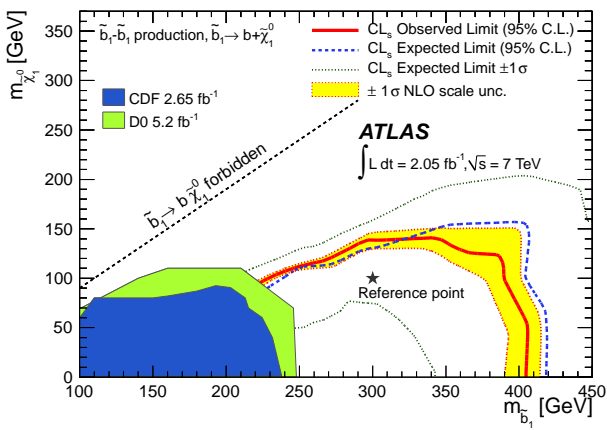}
    \end{center}
  \caption{Expected and observed exclusion limits, as well as $\pm 1 \sigma$ variation on the expected limit,  
  in the $\bone-\neut$ mass plane. The band around the observed limit delimited by the two dashed lines 
  shows the effect of renormalization and 
  factorization scale variation. The reference point indicated on the plane corresponds to the MSSM scenario with 
  sbottom and neutralino masses of 300~GeV and 100~GeV, respectively. 
  Results are compared to previous exclusion limits from Tevatron experiments. Results from LEP cover 
  the region with sbottom mass below 100~GeV. }
\label{fig:excl_sbLSP_dphi32}
\end{figure}
The results are translated into 95\% confidence-level (C.L.) upper limits on contributions from 
new physics using the CL$^{}_{\rm s}$ prescription~\cite{profilelike}. The SR with 
the best expected sensitivity at each point in parameter space is adopted as the 
nominal result. Figure~\ref{fig:excl_sbLSP_dphi32} shows the observed and 
expected exclusion limits at 95\% C.L. in the $\bone-\neut$ mass plane, assuming 
BR($\bone \rightarrow b\neut$)=100\%. 
Systematic uncertainties are treated as nuisance parameters and their correlations are 
taken into account. For the MSSM scenarios considered, the 
upper limit at 95\% C.L. on the sbottom masses obtained in the most conservative 
hypothesis, $\sigma^{}_{\rm{min}}$, is 390~GeV for $m^{}_{\neut} = 0$.
The limit becomes 405~GeV for $\sigma^{}_{\rm{nom}}$ 
and 420~GeV for $\sigma^{}_{\rm{max}}$. Neutralino masses of 120 GeV are 
excluded for $275<m_{\bone}<350$~GeV.   
The three signal regions are used to set limits on the effective cross 
section of new physics models, $\sigma^{}_{\rm eff}$, including the effects 
of experimental acceptance and efficiency. The observed (expected) excluded 
values of $\sigma^{}_{\rm eff}$ at 95\% C.L. are 13.4~fb, 9.6~fb and 5.6~fb 
(15.2~fb, 9.2~fb and 4.7~fb), respectively for \mct$>$100, 150, 200~GeV. 
%%% conclusions

In summary, we report results of a search for sbottom
pair production in $pp$ collisions at $\sqrt{s}$ = 7 TeV, based
on 2.05 fb$^{-1}$ of ATLAS data. The events are selected
with large \met~and two jets consistent with originating from 
\bquarks in the final state. The results are in agreement with SM
predictions for backgrounds and translate into 95\% C.L. 
upper limits on sbottom and neutralino masses in a given MSSM 
scenario for which the exclusive decay $\bone \rightarrow b\neut$ 
is assumed. For neutralino masses below 60~GeV, sbottom masses 
up to 390~GeV are excluded, significantly extending previous 
results. 

%%% acknoledgements
\section{Acknowledgements}

We thank CERN for the very successful operation of the LHC, as well as the
support staff from our institutions without whom ATLAS could not be
operated efficiently.

We acknowledge the support of ANPCyT, Argentina; YerPhI, Armenia; ARC,
Australia; BMWF, Austria; ANAS, Azerbaijan; SSTC, Belarus; CNPq and FAPESP,
Brazil; NSERC, NRC and CFI, Canada; CERN; CONICYT, Chile; CAS, MOST and
NSFC, China; COLCIENCIAS, Colombia; MSMT CR, MPO CR and VSC CR, Czech
Republic; DNRF, DNSRC and Lundbeck Foundation, Denmark; ARTEMIS, European
Union; IN2P3-CNRS, CEA-DSM/IRFU, France; GNAS, Georgia; BMBF, DFG, HGF, MPG
and AvH Foundation, Germany; GSRT, Greece; ISF, MINERVA, GIF, DIP and
Benoziyo Center, Israel; INFN, Italy; MEXT and JSPS, Japan; CNRST, Morocco;
FOM and NWO, Netherlands; RCN, Norway; MNiSW, Poland; GRICES and FCT,
Portugal; MERYS (MECTS), Romania; MES of Russia and ROSATOM, Russian
Federation; JINR; MSTD, Serbia; MSSR, Slovakia; ARRS and MVZT, Slovenia;
DST/NRF, South Africa; MICINN, Spain; SRC and Wallenberg Foundation,
Sweden; SER, SNSF and Cantons of Bern and Geneva, Switzerland; NSC, Taiwan;
TAEK, Turkey; STFC, the Royal Society and Leverhulme Trust, United Kingdom;
DOE and NSF, United States of America.

The crucial computing support from all WLCG partners is acknowledged
gratefully, in particular from CERN and the ATLAS Tier-1 facilities at
TRIUMF (Canada), NDGF (Denmark, Norway, Sweden), CC-IN2P3 (France),
KIT/GridKA (Germany), INFN-CNAF (Italy), NL-T1 (Netherlands), PIC (Spain),
ASGC (Taiwan), RAL (UK) and BNL (USA) and in the Tier-2 facilities
worldwide.

\bibliography{PUB_2j}

\begin{thebibliography}{45}
\bibitem [{atl()}]{atlascoordinates}%
  \BibitemOpen
  \href@noop {} {}\bibinfo {howpublished} {ATLAS uses a right-handed coordinate
  system with its origin at the nominal interaction point (IP) in the center of
  the detector and the $z$-axis coinciding with the axis of the beam pipe. The
  $x$-axis points from the IP to the center of the LHC ring, and the $y$-axis
  points upward. Cylindrical coordinates $(r,\phi)$ are used in the transverse
  plane, $\phi$ being the azimuthal angle around the beam pipe. The
  pseudorapidity is defined in terms of the polar angle $\theta$ as
  $\eta=-\ln\tan(\theta/2)$.}\BibitemShut {Stop}%
\end{thebibliography}
\onecolumngrid
\clearpage
\onecolumngrid
% ATLAS Collaboration author list for 02-NOV-2011
% Data extracted on 16-Mar-2012 for paperid 153
%\documentclass[11pt]{article}
%\usepackage{a4wide}\begin{document}
\begin{flushleft}
{\Large The ATLAS Collaboration}

\bigskip

G.~Aad$^{\rm 48}$,
B.~Abbott$^{\rm 110}$,
J.~Abdallah$^{\rm 11}$,
A.A.~Abdelalim$^{\rm 49}$,
A.~Abdesselam$^{\rm 117}$,
O.~Abdinov$^{\rm 10}$,
B.~Abi$^{\rm 111}$,
M.~Abolins$^{\rm 87}$,
O.S.~AbouZeid$^{\rm 157}$,
H.~Abramowicz$^{\rm 152}$
% ...
\bigskip

% ...
$^{10}$ Institute of Physics, Azerbaijan Academy of Sciences, Baku, Azerbaijan\\
$^{11}$ Institut de F\'isica d'Altes Energies and Departament de F\'isica de la Universitat Aut\`onoma  de Barcelona and ICREA, Barcelona, Spain\\
% ...
$^{48}$ Fakult\"{a}t f\"{u}r Mathematik und Physik, Albert-Ludwigs-Universit\"{a}t, Freiburg i.Br., Germany\\
$^{49}$ Section de Physique, Universit\'e de Gen\`eve, Geneva, Switzerland\\
% ...
$^{87}$ Department of Physics and Astronomy, Michigan State University, East Lansing MI, United States of America\\
% ...
$^{110}$ Homer L. Dodge Department of Physics and Astronomy, University of Oklahoma, Norman OK, United States of America\\
$^{111}$ Department of Physics, Oklahoma State University, Stillwater OK, United States of America\\
% ...
$^{117}$ Department of Physics, Oxford University, Oxford, United Kingdom\\
% ...
$^{152}$ Raymond and Beverly Sackler School of Physics and Astronomy, Tel Aviv University, Tel Aviv, Israel\\
% ...
$^{157}$ Department of Physics, University of Toronto, Toronto ON, Canada\\
% ...
\end{flushleft}

%\end{document}

\end{document}